\documentclass{PoS}

\title{Meson Electromagnetic Form Factors from Lattice QCD}

\ShortTitle{Meson Form Factors}

\author{\speaker{C. T. H. Davies}\\
        SUPA, School of Physics and Astronomy, University of Glasgow, Glasgow G12 8QQ, UK\\
        E-mail: \email{christine.davies@glasgow.ac.uk}}
\author{J. Koponen\\
        INFN, Sezione di Roma Tor Vergata, Via della Ricerca Scientifica 1, 00133 Roma, Italy\\}
\author{G. P. Lepage\\
        Laboratory for Elementary-Particle Physics, Cornell University, Ithaca, New York 14853, USA\\}
\author{A. T. Lytle\\
        INFN, Sezione di Roma Tor Vergata, Via della Ricerca Scientifica 1, 00133 Roma, Italy\\}
\author{A. C. Zimermmane-Santos\\
        S\~{a}o Carlos Institute of Physics, University of S\~{a}o Paulo, PO Box 369, 13560-1970 S\~{a}o Carlos, S\~{a}o Paulo, Brazil\\}
\author{HPQCD Collaboration\\
        http://www.physics.gla.ac.uk/HPQCD/}

\abstract{
Lattice QCD can provide a direct determination of meson electromagnetic form factors, 
making predictions for upcoming 
experiments at Jefferson Lab. The form factors are a reflection of the bound-state 
nature of the meson and so these calculations give information about how confinement 
by QCD affects meson internal structure. 
	The region of high squared (space-like) momentum-transfer, $Q^2$, is of particular 
interest because perturbative QCD predictions take a simple form in that limit that depends 
on the meson decay constant. 
We previously showed in~\cite{jonnaff} that, up to $Q^2$ of 6 $\mathrm{GeV}^2$, the 
form factor for a `pseudo-pion' made of strange quarks was significantly larger than 
the asymptotic perturbative QCD result and showed no sign of heading towards that value at higher $Q^2$. 
	Here we give predictions for real mesons, the $K^+$ and $K^0$, in anticipation of JLAB 
results for the $K^+$ in the next few years. 
	We also give results for a heavier meson, the $\eta_c$, up to $Q^2$ of 25 $\mathrm{GeV}^2$ for 
a comparison to perturbative QCD in a higher $Q^2$ regime. 
}

\FullConference{The 36th Annual International Symposium on Lattice Field Theory - LATTICE2018\\
		22-28 July, 2018\\
		Michigan State University, East Lansing, Michigan, USA.}

\begin{document}

\section{Introduction}

The determination of $\pi$ and $K$ electromagnetic form factors 
over a range of $Q^2$ values 
is a key set of experiments for the Jefferson Lab upgrade~\cite{jlab}. 
Since these form factors reflect the internal structure of the meson  
they test our understanding of the effects of the strong interaction 
if we can make accurate predictions for them from QCD. 
Lattice QCD enables us to do that, and we show results here for the $K$ meson 
in Section~\ref{sec:K}. 

A physical picture of the form factors at high $Q^2$ is provided by perturbative 
QCD (see Section~\ref{sec:pert}). 
This links a variety of exclusive processes together that can be factorised into 
meson `distribution amplitudes' combined with a hard scattering process. 
The meson distribution amplitudes derived from one process (or indeed calculated in lattice 
QCD) can be used in analysis of 
another, if perturbative QCD is a good description. 
As $Q^2 \rightarrow \infty$ this description becomes a particularly simple one 
(Eq.~\ref{eq:pert}) because the distribution amplitudes evolve with $Q^2$ to a simple form that is 
normalised by the decay constant. 
Testing this perturbative QCD picture 
has been hard because of the difficulty of obtaining accurate experimental 
information at sufficiently high $Q^2$. The upcoming Jefferson Lab experiments 
will provide important new information here. 

Lattice QCD can also provide new information to test perturbative QCD.  It is important to 
realise that, because this is a test of theory, this information does 
not have to relate to physical mesons that could be studied by experiment. 
We showed in~\cite{jonnaff} that lattice QCD calculations could be pushed to 
higher $Q^2$ than had been possible before by studying a `pseudopion' 
made of strange quarks. 
Here we go further, using charm quarks, and push up the determination of 
form factors to a $Q^2$ of 25$\mathrm{GeV}^2$, now in the genuinely high $Q^2$ regime.   

\section{Lattice QCD calculation}
\label{sec:latt}

\begin{figure}
\centering
\includegraphics[width=0.5\textwidth]{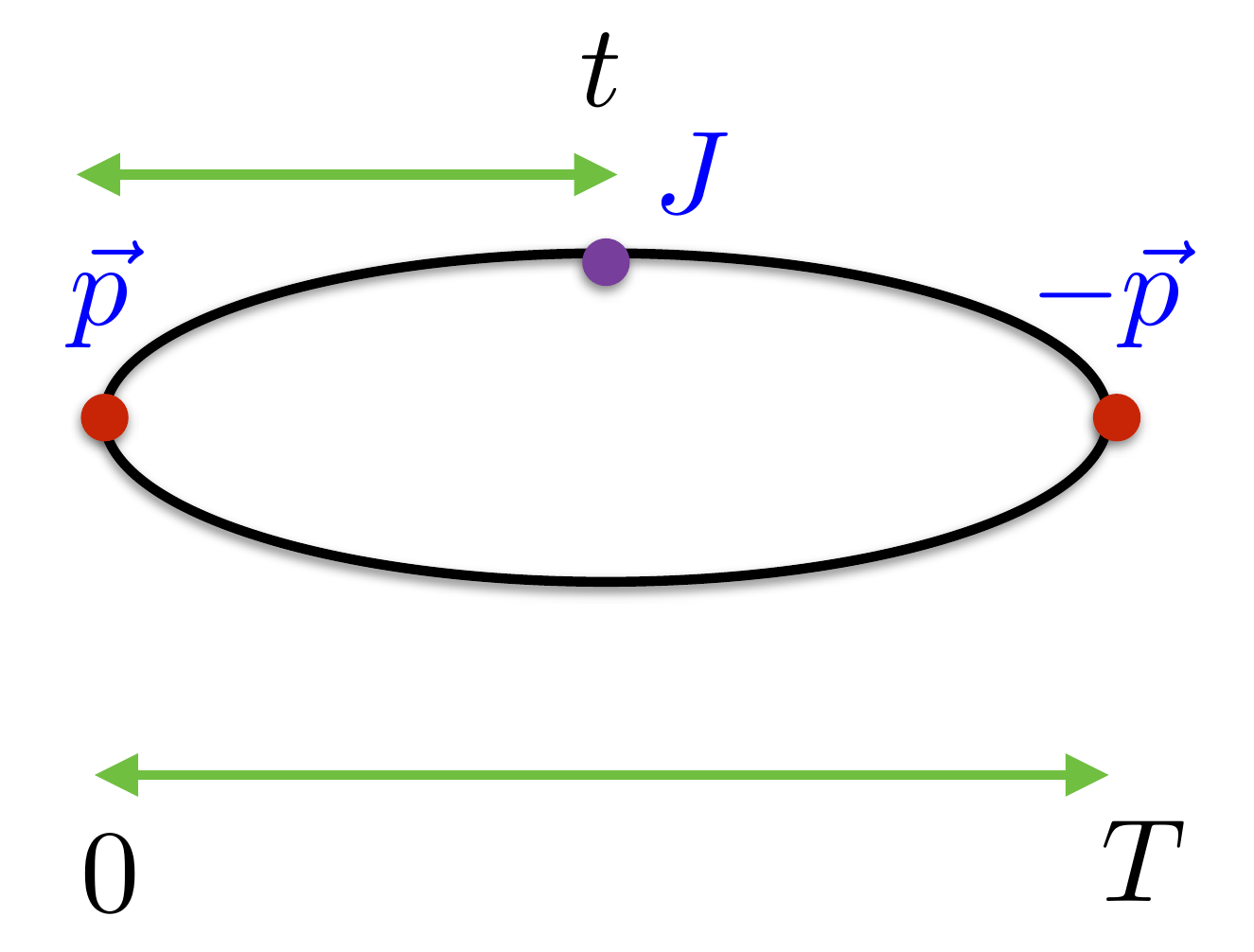}
\caption{Sketch of a 3-point correlator, showing the momentum configurations for the Breit frame (the spectator 
quark has zero spatial momentum) and the definition of $t$ and $T$.}
\label{fig:1}
\end{figure}

We use the Highly Improved Staggered Quark (HISQ) action~\cite{hisqdef} on high-statistics 
ensembles of gluon field configurations that include 2+1+1 flavours of HISQ 
quarks in the sea, generated by the MILC collaboration. 
For the $K$ meson results we use ensembles at 3 different values of the lattice 
spacing (0.15 fm, 0.12 fm and 0.09 fm approximately). 
The lattice spacing is determined by $w_0$, with the physical value of $w_0$ fixed from the 
pion decay constant~\cite{fkpi}. 
We have well-tuned valence $s$ quarks on each ensemble~\cite{mc}, and use valence light quarks with 
the same mass as the sea light quarks. The ensembles have sea light quarks 
with masses from 0.2 $\times$ that of the $s$ quark down to the physical point. 
We compare two different spatial volumes to test for finite-volume effects. 

	 For the $\eta_c$ results we use well-tuned valence $c$ quarks on ensembles with lattice 
spacing values of 0.09 fm, 0.06 and 0.045fm. 
These have sea light quarks with masses 0.2 $\times$ that of $s$ quarks only.  

	We calculate 2-point and 3-point correlation functions from quark propagators with 
zero and non-zero momentum. We use multiple time sources for increased statistics and insert spatial
momentum using twisted boundary conditions. 
For 3-point correlators (Figure~\ref{fig:1}) we use the Breit frame in which the initial and 
final states have equal and opposite spatial momentum. 
This maximises $Q^2$ for a given value of momentum in lattice units, $pa$. 

	The reach in $Q^2$ that is possible is limited more by the statistical errors 
that grow with $pa$ than by systematic errors at large $pa$ for our highly improved 
action~\cite{jonnaff}.  Access to higher $Q^2$ is possible on finer lattices. 

For the electromagnetic current, $J$, we use a one-link temporal vector current between `Goldstone' 
pseudoscalar mesons. For the $K$ meson we must calculate results for both light-quark and strange-quark currents. 
Simultaneous fits to 2-point and 3-point correlators as a function of $t$, for multiple $T$, at 
multiple momenta yield results for the matrix element 
\begin{equation}
\label{eq:me}
\langle P(\vec{p}) | J | P(-\vec{p}) \rangle = 2EF_{V,P}(Q^2) 
\end{equation}
with $Q^2 = |2\vec{p}|^2$.            
We normalise the current by dividing the form factor $F_V(Q^2)$ by its value at $Q^2=0$, 
where $F(0)=1$ by current conservation. 

We perform a similar calculation for the scalar current to extract a scalar form factor 
$F_{S,P}$ as a function of $Q^2$. In this case there are qualitative expectations for the 
shape of this form factor from perturbative QCD and so we also normalise that form 
factor by its value at $Q^2=0$ to study the shape at high $Q^2$.  

\section{Perturbative QCD expectation}
\label{sec:pert}

\begin{figure}
\centering
\includegraphics[width=0.8\textwidth]{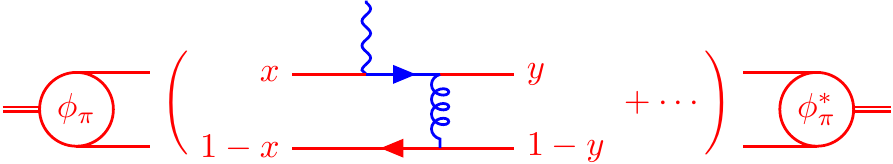}
\caption{The perturbative QCD description of the $\pi$ electromagnetic form 
factor at high $Q^2$. The blue lines indicate the route of high momentum transfer through the hard scattering process. }
\label{fig:2}
\end{figure}

The central hard-photon scattering factorises from 
the `distribution amplitudes', $\phi_P$, that describe the internal structure 
of the meson~\cite{lb}. See Figure~\ref{fig:2}. Redistribution of the photon momentum by 
gluons means that $F_V(Q^2)$ starts at $\mathcal{O}(\alpha_s)$ in a 
perturbative QCD approach. 
Normalisation of $\phi$ gives, at very high $Q^2$, :
\begin{equation}
\label{eq:pert}
Q^2 F_{V,P}(Q^2) = 8\pi\alpha_s f_P^2
\end{equation}
where $f_P$ is the decay constant of meson $P$. 
The scale of $\alpha_s$ is reasonably taken as $Q/2$ here~\cite{lb} 
because this is the momentum flowing through the gluon when 
the meson's quark and antiquark share its momentum equally. 
$\mathcal{O}(\alpha_s^2)$ corrections to the hard scattering process
have 
been calculated~\cite{rajan, melic} and have a coefficient 
of 1.18 when the scale 
of $\alpha_s$ is taken as $Q/2$. 

In the vector form factor case above, quark helicity is conserved at both 
the photon and the gluon vertices (up to quark mass effects) 
to allow a spin 0 meson to be turned around by the interaction. 
For the scalar form factor case, helicity would not be conserved at the 
scalar vertex and so we would expect the form factor to be suppressed by 
an additional power of $Q^2$ relative to the vector case.  

Note that these expectations, both qualitative and 
quantitative (Eq.~(\ref{eq:pert})), are predictions of theory and therefore hold
for pseudoscalar mesons made of quarks with unphysical masses as well 
as for physical ones. In lattice QCD 
we can calculate vector form factors for electrically neutral mesons  
by inserting a vector current on only one leg.  
Comparison of scalar form factors to expectations also now becomes possible. 
These examples show that the scope for testing perturbative QCD using fully 
nonperturbative lattice QCD results is not restricted 
to those mesons, or processes, which could be studied experimentally. 
Also note that no quark-line disconnected diagrams appear 
in Figure~\ref{fig:2} and so they are irrelevant to any comparison  
between lattice QCD and the perturbative QCD result. 

The key question to be answered is: at what $Q^2$ does perturbative physics start 
to be relevant to these form factors? Perturbative QCD allows us to 
connect these form factors, for example through determination of corrections to the 
distribution amplitudes, to other exclusive processes. If perturbative QCD is not valid 
until very high $Q^2$ values, then neither is this connection. 

\section{Results - $K$}
\label{sec:K}

\begin{figure}
\centering
\includegraphics[width=0.6\textwidth]{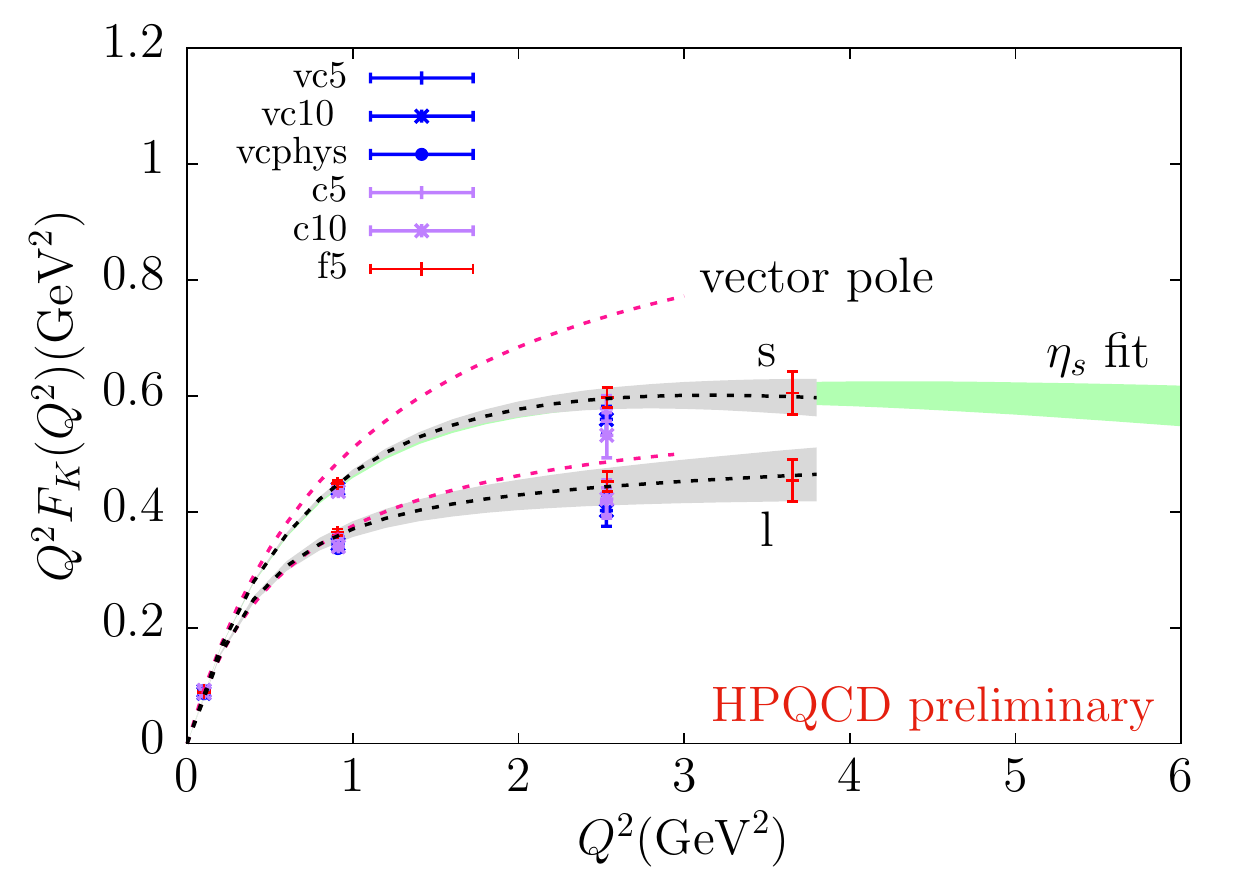}
\caption{The vector form factor for the $K$ meson. 
On the left we give results for the strange and light currents separately, with 
lattice results at three values of the lattice spacing and light quarks masses 
either $m_s/5$, $m_s/10$ or the physical value. The pink dotted lines show the form factor 
expected from vector pole-dominance. 
The fit results in the continuum limit for physical $u/d$ quark mass are shown as the grey 
bands.  The fit for the $s$-quark current is compared to our earlier 
result~\cite{jonnaff} for the $\eta_s$ in green. 
}
\label{fig:K1}
\end{figure}

\begin{figure}
\centering
\includegraphics[width=0.6\textwidth]{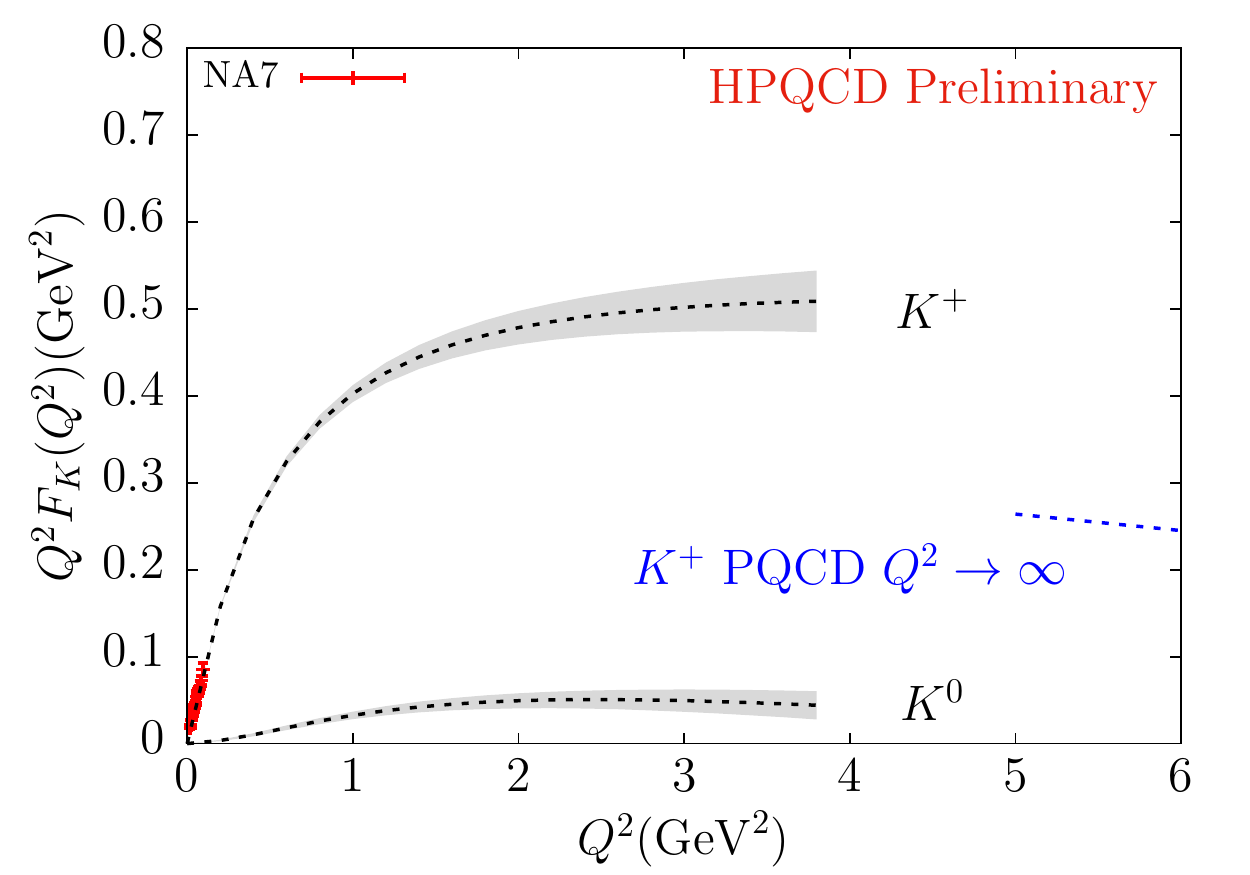}
\caption{The vector form factor for the $K^+$ and $K^0$ mesons. 
The fit results in the continuum limit for physical $u/d$ quark mass are shown as the grey 
bands. The red crosses show the results from NA7~\cite{na7} at small $Q^2$. 
}
\label{fig:K2}
\end{figure}

After fitting 2-point and 3-point correlators simultaneously to multi-exponential 
forms, we obtain the form factor at a range of $pa$ values on each gluon field ensemble studied. 
We then normalise the form factors as described and convert $pa$ values into $Q^2$ in $\mathrm{GeV}^2$ units using the lattice spacing. 

To interpolate in $Q^2$ and allow extrapolation to physical light quark masses and to 
$a=0$, we transform from $Q^2$ to $z$-space~\cite{jonnaff} and fit the combination 
$PF^q_{V,K}$ to a power-series in $z$, with coefficients that allow 
for $a$- and quark-mass dependence. 
Here $P$ is $(1+Q^2/M_v^2)$ where $M_v$ is the appropriate vector mass 
expected from pole-dominance of the form factor at low $Q^2$ 
(i.e. $\rho$ for the light-quark current and $F^l_{V,K}$ and $\phi$ for 
the $s$-quark current and $F^s_{V,K}$). 

	Figure~\ref{fig:K1} shows $K$ form-factor results separately for light- and 
strange-quark currents (with electric charge set to 1 in both cases). 
The grey band gives the continuum and chiral fit. $Q^2F_{V,K}$ is flat for both currents
above 2 $\mathrm{GeV}^2$. Note how the $s$-current result agrees with 
our earlier $\eta_s$ results~\cite{jonnaff}, even though the `spectator' quark is now a light one.

	Combining form factors with appropriate electric charge weights 
allows us to obtain form factors for $K^+$ and $K^0$ as in Figure~\ref{fig:K2}. 
Note the good agreement with NA7 results~\cite{na7} at low $Q^2$, 
but poor agreement with asymptotic perturbative QCD. 
There is also no sign of a trend downwards towards the perturbative 
result, as might be obtained from corrections to the asymptotic distribution amplitude. 

\section{Results - $\eta_c$}
\label{sec:etac}

\begin{figure}
\centering
\includegraphics[width=0.6\textwidth]{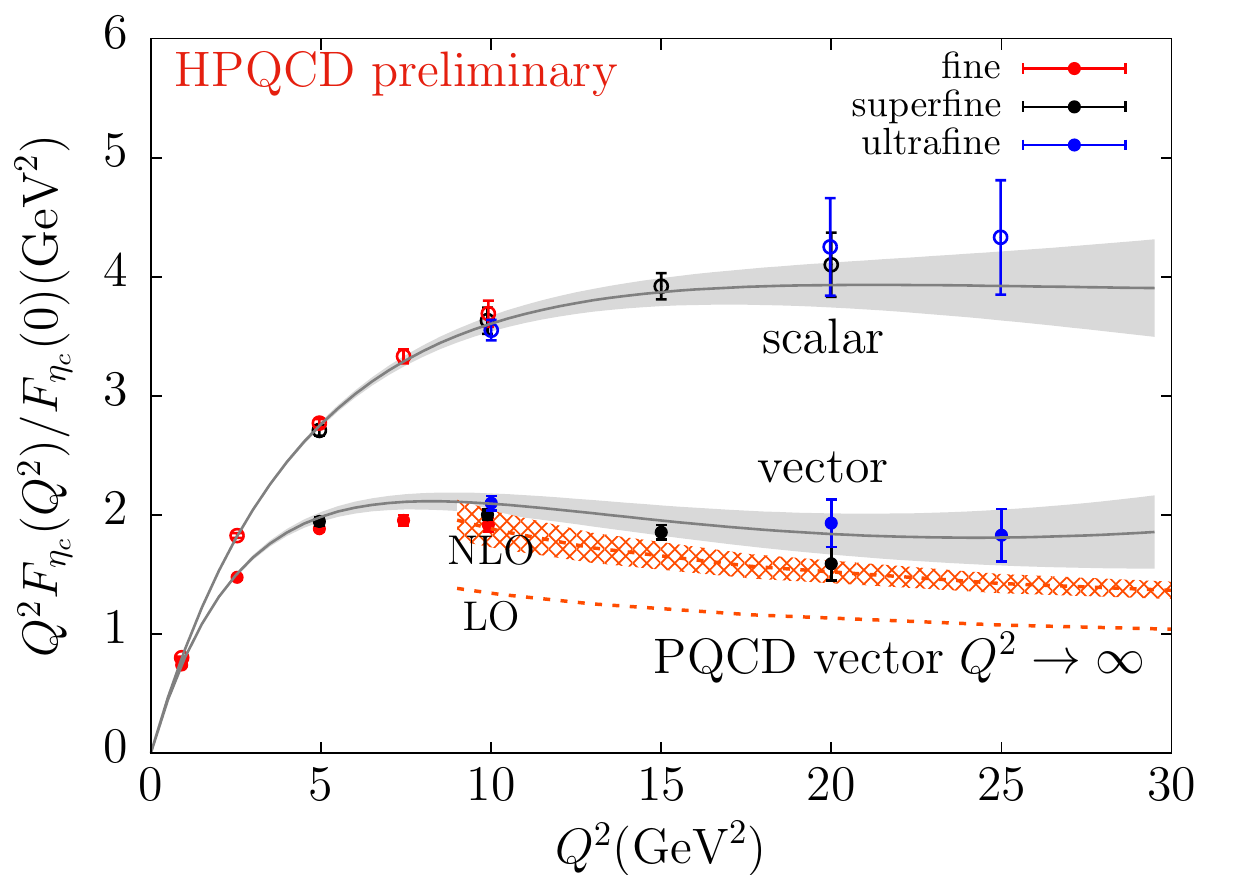}
\caption{The vector and scalar form factors for the $\eta_c$ meson 
(datapoints from 3 different values of the lattice spacing) 
and fit result extrapolated to $a=0$ (grey bands).
Both form factors are normalised by dividing by their value at $Q^2$ = 0. We also 
show the leading-order asymptotic perturbative QCD result (Eq.~(\ref{eq:pert}))
and the result from adding next-to-leading-order terms~\cite{rajan, melic} to the hard scattering kernel 
along with (hatched band) possible uncertainties from missing terms at $\alpha_s^3$.
}
\label{fig:etac}
\end{figure}

For valence charm quarks we are able to push to higher $Q^2$ values with good 
statistical precision. Figure~\ref{fig:etac} shows both the vector and scalar 
form factors (multiplied by $Q^2$) for the $\eta_c$, 
now up to $Q^2$ values well into what would normally 
be considered the perturbative regime. 

The vector form factor is closer to the perturbative QCD result than was true at lower 
mass. The shape of the scalar form factor, however, is similar to that of the vector and 
does not show any sign of falling faster than $1/Q^2$ (so that $Q^2\times F$ would fall)
as would be expected from the helicity arguments above~\cite{lb}. 

\section{Conclusions} 

We are able to calculate the electromagnetic form factor for the $K^+$ meson for the 
first time from lattice QCD. This gives a clear prediction, with uncertainties at 
the few percent level, for experiments starting at 
Jefferson Lab~\cite{jlab}.  We are able to obtain results up to $Q^2$ of $4\,\mathrm{GeV}^2$ 
and, although this is not yet a high value of $Q^2$, the disagreement with the asymptotic 
perturbative QCD result is substantial (a factor of 2). 

As further tests of perturbative QCD we can calculate both vector and scalar form factors 
for the pseudoscalar $\eta_c$ meson up to $Q^2$ = 25$\mathrm{GeV}^2$. 
Similar qualitative features to those in the $K$ case are seen, but with the gap closing somewhat 
between the vector form factor and the expected high-$Q^2$ perturbative QCD result. 

{\bf Acknowledgements} We are grateful to MILC for the use of their gluon field ensembles. 
This work was supported by the UK Science and Technology 
Facilities Council. The calculations used the DiRAC Data Analytic system at the University of 
Cambridge, operated by the University of Cambridge High Performance Computing Service 
on behalf of the STFC DiRAC HPC Facility (www.dirac.ac.uk). 
This is funded by BIS National e-infrastructure and 
STFC capital grants and STFC DiRAC operations grants.

\end{document}